\documentclass[prx,reprint]{revtex4-2}
\usepackage[normalem]{ulem}
\usepackage{graphicx}
\usepackage{amsmath}
\usepackage{amsfonts}
\usepackage{mathtools}
\usepackage{bm}
\usepackage[T1]{fontenc}
\usepackage{xfrac}

\newcommand{\ms}{m_s}
\newcommand{\mh}{m_h}
\newcommand{\mc}{m_c}
\newcommand{\mv}{m_v}
\newcommand{\bz}{B_z}

\begin{document}
\title{Weyl points and exceptional rings with polaritons in bulk semiconductors}
\author{R. L. Mc Guinness}
\author{P. R. Eastham}
\affiliation{School of Physics and CRANN, Trinity College Dublin, Dublin 2, Ireland.}

\date{\today}

\begin{abstract}
Weyl points are the simplest topologically protected degeneracy in a three-dimensional dispersion relation. The realization of Weyl semimetals in photonic crystals has allowed these singularities and their consequences to be explored with electromagnetic waves. However, it is difficult to achieve nonlinearities in such systems. One promising approach is to use the strong coupling of photons and excitons, because the resulting polaritons interact through their exciton component. Yet topological polaritons have only been realized in two dimensions. Here, we predict that the dispersion relation for polaritons in three dimensions, in a bulk semiconductor with an applied magnetic field, contains Weyl points and Weyl line nodes. We show that absorption converts these Weyl points to Weyl exceptional rings. We conclude that bulk semiconductors are a promising system in which to investigate topological photonics in three dimensions, and the effects of dissipation, gain, and nonlinearity.
\end{abstract}


\maketitle

\section{Introduction}

Degeneracies in bandstructures are a key concept at the heart of
recent developments in condensed-matter physics and
optics \cite{ozawa_topological_2019}. Two-dimensional materials such as
graphene possess Dirac points, where the dispersion is locally linear,
which are responsible for many of their unique properties. In
three-dimensional materials Weyl points have been found in
photonic \cite{ExpObsWeyl} and electronic \cite{xu_discovery_2015}
bandstructures, providing low-energy models of Weyl fermions. More
generally, topological considerations mean that materials hosting
degeneracies are the basis for realizing topological insulators and
related effects such as robust edge modes. Such work is now also being
extended to dissipative systems, such as photonic materials with gain
and loss, described by non-Hermitian Hamiltonians \cite{NHTop}. In this
case the singularities include exceptional
points \cite{NHandPT,heiss_physics_2012}
in parameter space, at which both the frequencies and lifetimes of the
modes become degenerate. Rings of such exceptional points have been shown to emerge from Dirac points in photonic crystals \cite{EPRing}. In the three-dimensional case, Weyl points can become Weyl exceptional rings \cite{xu_weyl_2017}, which have a quantized Chern number and a quantized Berry phase. Like their counterparts in Hermitian systems, such non-Hermitian singularities give rise to  interesting
physical effects \cite{RotExc}, including edge
modes \cite{leykam_edge_2017}, unusual transmission properties,
topological lasing, and Fermi arcs arising from half-integer
topological charge \cite{HalfPol}.

Polaritons are exciton-photon superpositions that are formed by strong
light-matter coupling in semiconductors\
\cite{kavokin_microcavities_2017,YuCard}. Their
half-matter half-light nature implies relatively strong
nonlinearities, and this feature among others makes them an
interesting system in which to study topological effects. Topological
phases have been
predicted \cite{ge_floquet_2018,PolZG,PolZ,li_lieb_2018,YiKar,TopPol} and
observed \cite{klembt_exciton-polariton_2018} for polaritons formed
from quantum-well excitons coupled to photons confined in
microcavities. Topological lasing \cite{kartashov_two-dimensional_2019}
and exceptional points \cite{EPPol} have also been studied. However, as
microcavities and quantum-wells are two-dimensional systems, phenomena
such as Weyl points, Fermi arcs, and the three-dimensional topological
phases \cite{ozawa_topological_2019}, have not been considered.

In this paper, we report topologically nontrivial dispersion
relations for polaritons propagating in three dimensions. We consider
a bulk semiconductor in a magnetic field and show that the p-type
structure of the valence band leads to intricate dispersion relations containing topologically protected degeneracies. In the absence of nonradiative
losses there are eight sheets of the dispersion surface, which host
Weyl points \cite{ExpObsWeyl, xu_discovery_2015, WeylAc, WeylObs2,
  yang_ideal_2018}, for wave vectors along the
field direction, and ring degeneracies, for
wave vectors transverse to it. In the non-Hermitian
case \cite{NHTop,leykam_edge_2017,NHandPT}, with absorption, we show that the Weyl points become Weyl exceptional rings, which can be
reached by tuning the frequency and the angle between the propagation
direction and the applied field.  These results show that bulk
semiconductors could be used to study topological effects in three
spatial dimensions. Furthermore, bulk polariton lifetimes can be long
since, unlike microcavity polaritons, they are not subject to
radiative decay.  They may therefore give access to the
strongly interacting regime of topological
photonics \cite{ozawa_topological_2019}.

\section{Method}

\subsection{Exciton spectra}

We consider polaritons formed from 1s excitons in direct band-gap zinc-blende
semiconductors such as GaAs. These involve p-type valence band states
with $\Gamma_8$ symmetry, and s-type conduction band states with
$\Gamma_6$ symmetry. The combinations of the hole spin $\mh=\pm
\sfrac{3}{2},\pm \sfrac{1}{2}$ and the electron spin $\ms=\pm \sfrac{1}{2}$ then give rise to
eight exciton spin states, denoted $|X_{n}\rangle$ for $n=1,\ldots 8$, with energies $E_{n}$. 

\begin{figure}[t]
\includegraphics{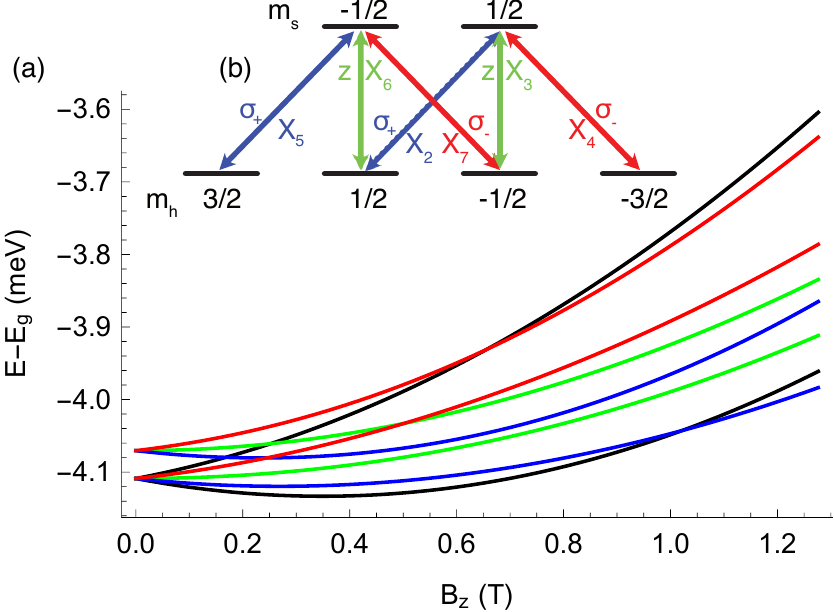}
\caption{(a) Calculated 1s exciton energies relative to the bandgap
  for GaAs with a magnetic field $\bz \in [0,B_{\mathrm{max}}]$ in the $[001]$ direction. The
  line coloring indicates the polarization of each transition:
  right circular (red/$\sigma_-$) and left circular (blue/$\sigma_+$),
  with the field in the xy plane, or linear in the z direction
  (green/z). The black curves are the spin-2 dark
  excitons. (b) Polarization and spin structure of the exciton
  transitions, in terms of the hole spin $\mh$ and electron spin $\ms$.
}
\label{Fig:1sexcitons}
\end{figure}

To evaluate the polariton spectrum, we need the energies and
polarizations of the exciton transitions. To obtain these, we
diagonalize the effective Hamiltonian for the 1s excitons given in
Ref.\ \cite{1sXCho}. The parameters in this effective
Hamiltonian are related to the underlying electron-hole exchange
parameters, Luttinger parameters, and g-factors. This approach treats
the valence-band anisotropy, magnetic field, and electron-hole
exchange as perturbations on a spherically-symmetric electron-hole
Hamiltonian~\cite{AltLipB}. The unperturbed wavefunction is of the
usual hydrogenic form, with the binding energy
$R_0=\mu e^4/32\pi^2\epsilon_0^2\epsilon^2\hbar^2$ and Bohr radius
$a_0=4\pi\epsilon_0\epsilon\hbar^2/\mu e^2$, where
$\mu^{-1}=\mc^{-1}+\mv^{-1}$. $\mv=m_{0}/\gamma_1$ is the isotropic
part of the effective mass for the valence band, related to the
Luttinger parameter $\gamma_1$, and $\mc$ is the effective mass for
the conduction band. For this perturbative approach to be valid the
cyclotron energy must be small compared with the exciton binding
energy $R_0$. We take the specific criterion given by Altarelli and
Lipari~\cite{AltLipB}, \begin{equation}
  \gamma=\frac{\hbar\omega_c}{2R_0}=\frac{e\hbar B}{2\mu R_0}\leq
  0.4,\end{equation} to define the maximum field $B_{\mathrm{max}}$ of
the perturbative regime. In the following we will consider the
specific case of GaAs, with applied field $B_{\mathrm{max}}$ in the
$[001]$ direction, using the bandstructure parameters from
Ref.~\cite{Wink}. For the electron-hole exchange
parameters~\cite{1sXCho} we take
$\Delta_1=-9.61\ \mathrm{\mu eV}$~\cite{fu_excitonic_1999}, and
$\Delta_0=\Delta_2=0$.  The exciton spectrum computed for these
parameters is shown in Fig.~\ref{Fig:1sexcitons}(a). As expected, the
magnetic field lifts the degeneracies of the eight electron-hole pair
states. This splitting of the energies of the excitons will result in
an anisotropic and multiply resonant optical susceptibility and hence
a direction- and polarization-dependent polariton dispersion.
 
\subsection{Polariton Hamiltonian}

The topological singularities of the polariton dispersion arise from
the polarization dependence of the exciton-photon coupling. In the
Coulomb gauge the interaction between the vector potential and the
electrons, from the Hamiltonian
$\sum_i [\hat{\bf{p}}_i+e\hat{\bf{A}}(\hat{\bf{r}}_i)]^2/(2m)$,
is \begin{equation}
  \hat{H}_{ep}=\frac{e}{m}\sum_i\sum_{\mathbf{k},s}\sqrt{\frac{\hbar}{2\epsilon_0\omega
      V}}\left[\hat{a}_{{\bf{k}},s}\mathbf{e}_{\mathbf{k},s}e^{i\mathbf{k}.\hat{\mathbf{r}}_i}+\mathrm{H.\;c.}\right]\cdot\hat{\bf{p}}_i.\end{equation}
Here, the first sum is over the electrons, and the second is over the
photon wave vectors, $\bf{k}$, and polarizations, $s$, with
corresponding polarization vectors
$\mathbf{e}_{\mathbf{k},s}$. $\hat{a}_{\mathbf{k},s}$ is the photon
annihilation operator, $\omega=c|{\bf{k}}|$ is the photon frequency, and
$V$ is a quantization volume. Thus we have the second-quantized
Hamiltonian in the subspace of the eight 1s exciton states,
$|X_{\mathbf{k},n}\rangle$, \begin{multline}
  \hat{H}_{xp}=\frac{e}{m}\sum_{\mathbf{k},s,n}
  \sqrt{\frac{\hbar}{2\epsilon_0\omega
      V}}\big[\hat{X}^{\dagger}_{\mathbf{k},n}\hat{a}_{\mathbf{k},s}\mathbf{e}_{\mathbf{k},s}\cdot\langle
  X_{\mathbf{k},n}|\hat{\mathbf{p}}|0\rangle \\ +
  \mathrm{H.c.}\big],\label{eq:secondquantham}\end{multline} where we
have made the rotating-wave approximation. We have also neglected the diamagnetic term, proportional to $\hat{\bm{A}}^2$, which is justified when treating the strong coupling of near-resonant modes \cite{schafer_relevance_2020}. Both the diamagnetic and counterrotating terms give nonresonant contributions which are perturbative in the ratio of the light-matter coupling strength to the exciton or photon energy, and hence small. There is, also, a resonant part of the diamagnetic term, which is accounted for in the observed value of the exciton energy. We note that the diamagnetic and counterrotating terms can be important away from resonance, being required, for example, to ensure the photon energy remains positive at long wavelengths. They are also important in the ultra-strong coupling regime, where the light-matter coupling becomes comparable to the transition energy.

In the envelope function approximation the matrix
elements appearing in Eq. (\ref{eq:secondquantham}) are products of
the matrix elements of the Bloch functions at $\mathbf{k}=0$ and the
hydrogenic exciton wavefunctions $\chi_{m_s,m_h}F_{n}(\mathbf{r}=0)$. For the spatial part of the latter we take the unperturbed result $|F_n(0)|^2\approx1/\pi a_0^3$. For the spin part $\chi_{m_s,m_h}$ we note that at the field $B_{\mathrm{max}}$ we are considering the Zeeman terms dominate over
the electron-hole exchange. Thus the excitons are, to a good approximation, diagonal in the spin projections $m_s$ and $m_h$\ \footnote{Electron-hole exchange mixes different electron-hole spin configurations within each polarization, i.e., total spin $m_s+m_h$, modifying the oscillator strength and hence producing small changes in the spectrum. At the field $B_{max}$ the dominant component corresponds to between 93\% and 99\% of the wavefunction, depending on the exciton.}. Using the standard forms
for the valence-band wave functions \cite{XCho} and the Kane
parameter $P$ \citep{Basu,Sem3,SwierNC}, we then have $\langle X_{n}|\hat{\mathbf{p}}|0\rangle=\sqrt{V} mF^{\ast}_n(0)|P|\mathbf{v}_n/\hbar$, where $\mathbf{v}_1=\mathbf{v}_8=\mathbf{0}$ and 
\begin{align}
\mathbf{v}_2 &=\frac{1}{\sqrt{6}}\begin{pmatrix}i \\ 1 \\ 0\end{pmatrix}  & \mathbf{v}_3 &=\sqrt{\frac{2}{3}}\begin{pmatrix} 0 \\ 0 \\ i\end{pmatrix} &
\mathbf{v}_4 &=\frac{1}{\sqrt{2}}\begin{pmatrix}-i \\ 1 \\ 0\end{pmatrix}
\nonumber \\ 
\mathbf{v}_5 &=\frac{1}{\sqrt{2}}\begin{pmatrix}i \\ 1 \\ 0\end{pmatrix} & \mathbf{v}_6 &=\sqrt{\frac{2}{3}}\begin{pmatrix}0 \\ 0 \\ i\end{pmatrix} &
\mathbf{v}_7 &=\frac{1}{\sqrt{6}}\begin{pmatrix}-i\\ 1 \\ 0\end{pmatrix}
.\label{eq:matelems}
\end{align}
The states $|X_1\rangle\ldots|X_4\rangle$ correspond to excitons with electron spin $\ms=\sfrac{1}{2}$ and hole spin $\mh=\sfrac{3}{2},\sfrac{1}{2},-\sfrac{1}{2},-\sfrac{3}{2}$, respectively, while $|X_5\rangle\ldots|X_8\rangle$ are the corresponding states with $\ms=-\sfrac{1}{2}$. Thus $|X_1\rangle$ and $|X_8\rangle$ are dark states, and the remaining transitions are either circularly polarized in the xy plane or linearly polarized in the z direction, as shown in Figs.~\ref{Fig:1sexcitons}(a) and ~\ref{Fig:1sexcitons}(b).

Using these exciton matrix elements in Eq. (\ref{eq:secondquantham}), and approximating $\omega\approx E_g/\hbar$ in the prefactor of the coupling, we find for the exciton-photon Hamiltonian, \begin{align}
\hat{H}=\sum_{\mathbf{k}}&\bigg\{ \sum_{s}\frac{\hbar c k}{\sqrt{\epsilon}} \hat{a}^\dagger_{\mathbf{k},s}\hat{a}_{\mathbf{k},s} + \sum_{n} E_{n} \hat{X}^{\dagger}_{\mathbf{k},n}\hat{X}_{\mathbf{k},n} \nonumber \\  & + \frac{\hbar\Omega_0}{2}\sum_{s,n} (\mathbf{e}_{\mathbf{k},s}\cdot \mathbf{v}_{n})\hat{X}^{\dagger}_{\mathbf{k},n}\hat{a}_{\mathbf{k},s} + \mathrm{H.\;c.}\bigg\},\label{eq:exphotham}\end{align} where $k=|\mathbf{k}|$ and we use a Rabi splitting \begin{displaymath}\hbar\Omega_0=\sqrt{\frac{2 e^2 |P|^2}{\epsilon_0\epsilon E_g \pi a_0^3}}\end{displaymath} to quantify the light-matter coupling in the material.

\subsection{Polariton spectra}

In the following we will consider the polariton spectrum, which we
obtain from the Heisenberg equations of motion for the exciton and
photon annihilation operators by looking for solutions with time
dependence $e^{-i\omega t}$, i.e., setting
$\hat{a}(t)=e^{-i\omega t}\hat{a}(0)$ and similarly for the exciton
operators. We specify the wavevector direction in terms of the polar
coordinates $\theta,\phi$, with the field direction and the $[001]$
crystal axis corresponding to $\theta=0$.  For the photon polarization
we use the circularly polarized states,
$\mathbf{e}_{\mathbf{k},\pm}=(\mathbf{e}_{\mathbf{k},\theta}\pm
i\mathbf{e}_{\mathbf{k},\phi})/\sqrt{2}$, constructed from the
linearly polarized basis transverse to $\mathbf{k}$,
\begin{align}
\mathbf{e}_{\mathbf{k},\theta}&=\begin{pmatrix} \cos\theta \cos\phi \\ \cos\theta
\sin\phi \\ -\sin\theta\end{pmatrix}, & \mathbf{e}_{\mathbf{k},\phi}&=\begin{pmatrix} - \sin\phi
 \\ \cos\phi \\ 0\end{pmatrix}.\label{eq:polvecslin}\end{align}
Since there are two photon polarizations and six excitons (discounting the irrelevant dark states), this procedure gives an $8\times 8$ Hamiltonian matrix, $H_8$, with elements dependent on the wavevector magnitude and direction. The form of this Hamiltonian is given explicitly in the Appendix. We determine the polariton dispersion $\omega(k,\theta,\phi)$ by finding the eigenvalues, $\omega$, of $H_8$ numerically. 

While the dispersion $\omega(k,\theta,\phi)$ is given by the
eigenvalues of $H_8$, there is another approach to analyzing the
topological singularities of the polariton spectrum, in terms of the
function $k(\omega,\theta,\phi)$. This latter function provides a
natural description of optics at a fixed frequency and is related to
constructs such as the refractive index surface of classical
optics \cite{BornWolf,BerryBr1,ballantine_conical_2014}. For example, a radial plot of $k$ at some fixed frequency over
angles gives a contour (in $k$ space) of the dispersion relation
$\omega(k)$. The normal to such an isofrequency surface is therefore
the group velocity, $\bm{\nabla}_\mathbf{k} \omega$, controlling the refraction
direction at that frequency. While the two functions
$\omega(k,\theta,\phi)$ and $k(\omega,\theta,\phi)$ are equivalent in
the absence of dissipation, we shall see that they have some
differences in its presence, and we therefore consider both
representations in the following.

To obtain the dispersion in the form $k(\omega,\theta,\phi)$, we eliminate the exciton amplitudes from the Heisenberg equations of motion. This leads to a two-dimensional eigenproblem for the photon amplitudes, \begin{align}\bigg[&(\omega-ck/\sqrt{\epsilon})\delta_{ss^{'}} \nonumber \\ & -\frac{\Omega_0^2}{4}\sum_{n}\frac{(\mathbf{e}^\ast_{\mathbf{k},s}\cdot\mathbf{v}^\ast_n).(\mathbf{e}_{\mathbf{k},s^{'}}\cdot\mathbf{v}_n)}{\omega-E_n/\hbar}\bigg]\hat{a}_{\mathbf{k},s'}(0)=0,\label{eq:twobytwomatrix}\end{align} so that the magnitudes of the wavevectors are the eigenvalues of a $2\times 2$ matrix, whose elements are functions of the frequency and propagation direction. 

It is useful to note that this form, Eq. (\ref{eq:twobytwomatrix}), can also be derived semiclassically, by looking  for plane-wave solutions to Maxwell's equations, including the excitonic resonances
via a frequency-dependent dielectric function $\epsilon(\omega)$. In an optically isotropic material, 
$\epsilon(\omega)$ is a scalar, and the polariton dispersion satisfies
$c^2 k^2/\omega^2=\epsilon(\omega)$ \cite{HuagKoch}. In the present case, however, the optical response is anisotropic due to the magnetic field, and we must consider the vector equation
\begin{equation}
-\operatorname{\mathbf{k}\times \mathbf{k} \times} \mathbf{E} = \operatorname{\frac{\omega^2}{c^2} \epsilon(\omega)} \mathbf{E}. \label{eq:PolS}
\end{equation} Longitudinal modes, with $\mathbf{k}\parallel \mathbf{E}$, occur if $\epsilon(\omega)=0$. To obtain the equation for the transverse modes, we take matrix elements of Eq. \eqref{eq:PolS} in a basis perpendicular to $\hat{\mathbf{k}}$, such as $\mathbf{e}_{\mathbf{k},\pm}$. This eliminates the zero eigenvalue of the operator
$\mathbf{k}\times \mathbf{k}\times$, i.e. the longitudinal polariton,
and gives
\begin{equation}
\left[(\omega^2-c^2k^2/\epsilon)\delta_{ss^\prime}+\frac{\omega^2}{\epsilon}\chi_{ss^\prime}\right]\mathbf{E}_{\mathbf{k},s^\prime}=0,\end{equation} where $\chi_{ss^\prime}(\omega)=\mathbf{e}^\dagger_{\mathbf{k},s}\chi(\omega)\mathbf{e}_{\mathbf{k},s^\prime}$
is the transverse part of the excitonic susceptibility and $\epsilon$ is the background permittivity. We approximate the prefactors in this expression as $(\omega^2-c^2k^2/\epsilon)\approx 2\omega(\omega-ck/\sqrt{\epsilon})$ and $\omega^2/\epsilon\approx\omega E_g/\hbar\epsilon$. Comparing this expression with Eq. (\ref{eq:twobytwomatrix}), we see that the final term in the latter is related to the susceptibility by\begin{equation} \chi_{ss^\prime}=-\frac{\Omega_0^2\hbar\epsilon}{2E_g}\sum_{n}\frac{(\mathbf{e}^\ast_{\mathbf{k},s}\cdot\mathbf{v}^\ast_n).(\mathbf{e}_{\mathbf{k},s^{'}}\cdot\mathbf{v}_n)}{\omega-E_n/\hbar}.\end{equation}

The spectrum $k(\omega,\theta,\phi)$ can be found straightforwardly by
solving the secular equation for Eq. (\ref{eq:twobytwomatrix}), which
is a quadratic in $k$. It may be noted that $\chi_{ss^\prime}$, and
hence the polariton spectrum, is independent of $\phi$. This reflects
the rotational symmetry of the problem about the magnetic field
($\theta=0$). The combination of the form of
Eq. \eqref{eq:twobytwomatrix} with that of $\chi_{ss^\prime}$
imposes an additional symmetry between the solutions at $\theta$ and
those at $\pi - \theta$. We may therefore set $\phi=0$ and consider
the interval $\theta\in [0,\frac{\pi}{2} ]$. We note that whereas the
secular equation for Eq. (\ref{eq:twobytwomatrix}) is a quadratic in
$k$, that for $H_8$ is an eighth-order polynomial in $\omega$. Thus
there are, in general, two wave vectors for each frequency, from the
two dispersing photon modes. There are, however, eight frequencies for
each wave vector, from those two photon modes as well as the six
nondispersing bright excitons.

\section{Results}

\subsection{Topological singularities: Hermitian case}

Our primary interest is in the degeneracy structure of the
magneto-exciton-polariton dispersion relation, which we first consider
in the Hermitian case without dissipation. It is possible to make some
observations that constrain the possible degeneracies based on the
symmetry of the problem. Owing to the $\phi$ invariance of the
solutions, we know that propagation in the $\theta=0$ direction is the
only configuration for which isolated degeneracies are
possible. Correspondingly, if degeneracies occur at any non-zero
$\theta$, they are necessarily extended degeneracies over all $\phi$.

\begin{figure}
\includegraphics{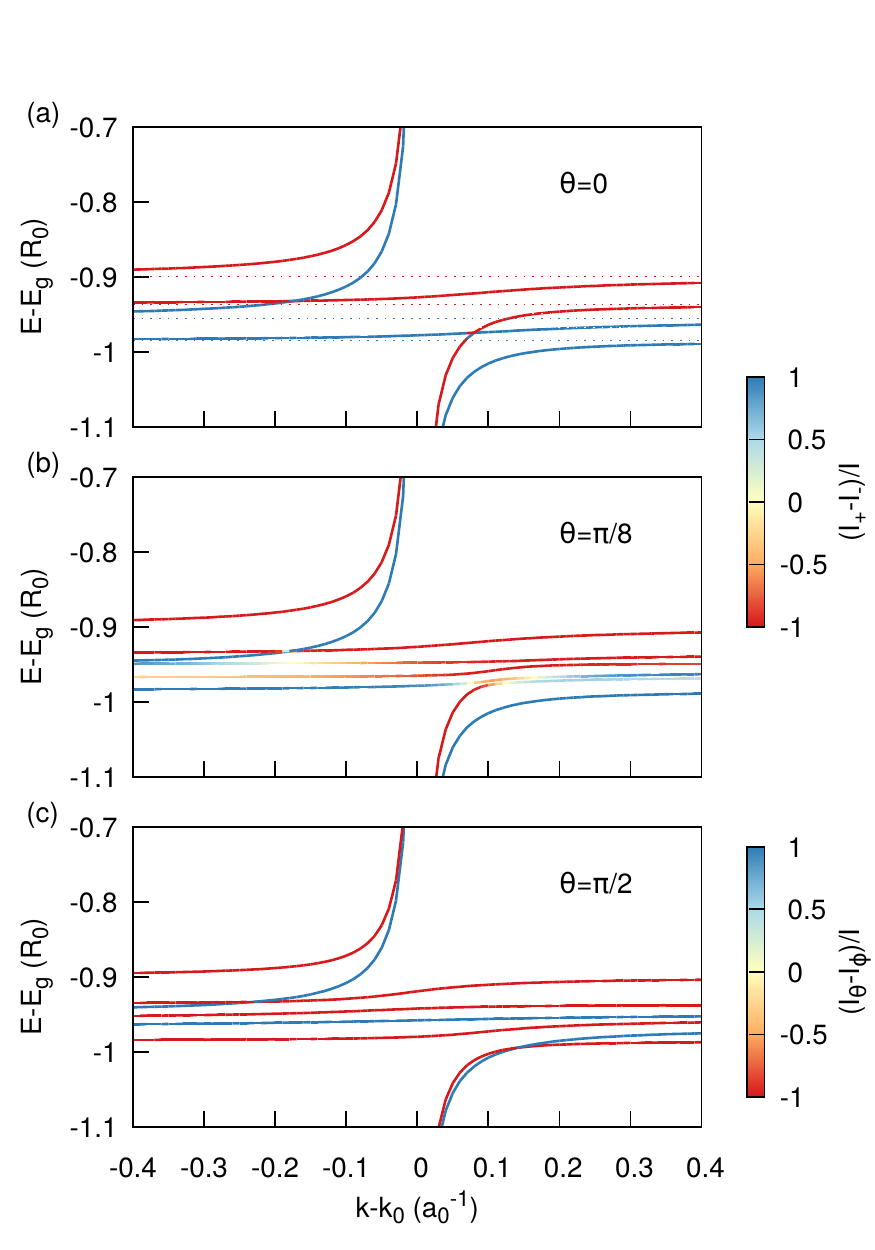}
\caption{(a) Solid curves: polariton dispersion relations for GaAs in a magnetic field for $\theta=0$. The coloring shows the degree of circular polarization. Dotted lines: energies of the z-polarized excitons. (b) Polariton dispersion for $\theta=\pi/8$. (c) Polariton dispersion for $\theta=\pi/2$. The coloring here shows the degree of linear polarization. Energies are relative to the band gap in units of the exciton Rydberg, and wave vectors are relative to $k_0$ in units of the inverse exciton Bohr radius.}
\label{Fig:disps}
\end{figure}

In Fig.\ \ref{Fig:disps} we plot the dispersion of the transverse
modes, obtained from the secular equation for
Eq. (\ref{eq:twobytwomatrix}). The three panels refer to propagation
along the field, $\theta=0$, at a small angle to it, $\theta=\pi/8$,
and perpendicular to it, $\theta=\pi/2$. The polarization of the modes
is shown by the coloring. Energy is measured relative to the band gap
$E_g$ and in units of the exciton Rydberg energy. The wave vector is
measured relative to $k_0=\sqrt{\epsilon}(E_g-R_0)/\hbar c$, which is
the wave vector at which the bare linear photonic dispersion would
cross an unperturbed exciton. The wave vector is measured in units of
the inverse exciton Bohr radius.

The spectrum for propagation along the z-axis is shown in
Fig. \ref{Fig:disps}(a). In this case the two z-polarized excitons,
$X_{3}$ and $X_{6}$, do not couple to light, and there are only six
modes in the transverse spectrum. The other excitons are circularly
polarized, and each circular polarization of light mixes with the two
excitons of that polarization. This gives rise to a spectrum with a
lower, intermediate, and upper branch for each circular
polarization. The lower branch begins at low energy as a purely
photonic, linearly dispersing, mode, which anticrosses with the lower-energy exciton, asymptoting horizontally to approach that exciton
energy at large $k$.  Above that energy there is an intermediate
branch, which initially has an imaginary $k$ as it lies in the
polaritonic (longitudinal-transverse) gap\
\cite{YuCard}. This mode then becomes a propagating
polariton with $k=0$ at the gap edge and then approaches the higher
exciton energy at large $k$. Above this there is an upper branch,
which again begins as a solution with imaginary $k$, before becoming a
propagating solution, and finally approaching the photon dispersion at
large $k$.

Figure\ \ref{Fig:disps}(b) shows the spectrum at a small angle to the
z-axis. Comparing this spectrum with that in Fig.\ \ref{Fig:disps}(a) we
see that there are now eight branches, because the z-polarized
excitons now couple to light. Moreover, we find that this spectrum is
non-degenerate, with avoided crossings which originate from the degeneracies at
$\theta=0$. The two degeneracies in Fig.\ \ref{Fig:disps}(a) between the
different circular modes have split. The splittings are quite small, due to the small angle chosen, but are nonetheless present. In addition, we see that the three
intersections between the z-polarized excitons and the transverse
modes at $\theta=0$ have split, to organize into the two additional
transverse branches at $\theta\neq 0$. There is in fact a fourth
intersection of this nature, involving the highest energy z-polarized
exciton, but it is with an evanescent mode at an imaginary $k$.

\begin{figure}
\includegraphics{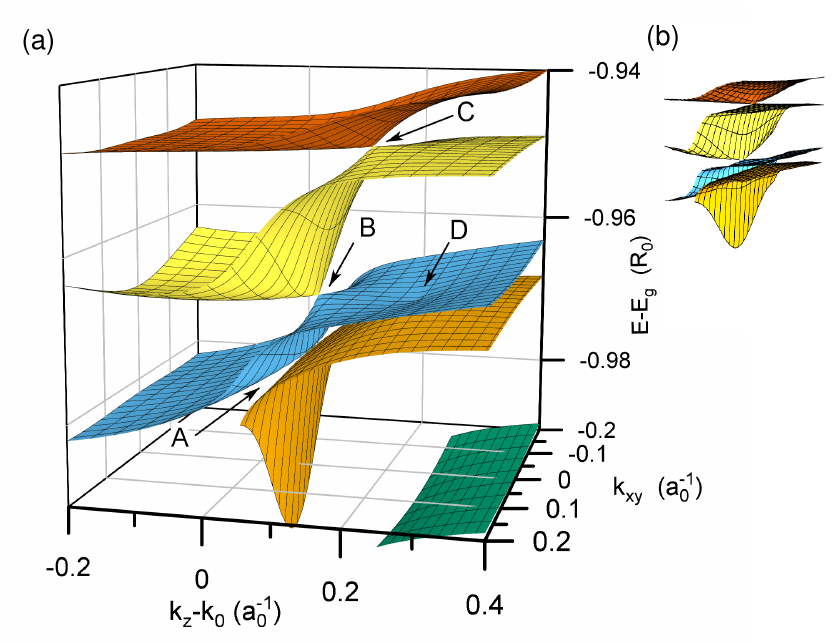}
\caption{(a) Polariton dispersion relation for GaAs in a magnetic field, centered around propagation along the magnetic-field
  direction, z. The dispersion relation has rotational symmetry around the $z$ axis. It is shown as a function of $k_z$ and the wavevector in the xy plane, $k_{xy}=\pm\sqrt{k_x^2+k_y^2}$. Two Weyl points with a linear dispersion in all directions, B and C, and a band touching with a mixed quadratic/linear
  dispersion, A, can be seen. A third Weyl point, D, lies between the orange and blue sheets (second and third from the bottom). (b) The same data from a different perspective.}
\label{Fig:Weyl2}
\end{figure}

\begin{figure}
\includegraphics{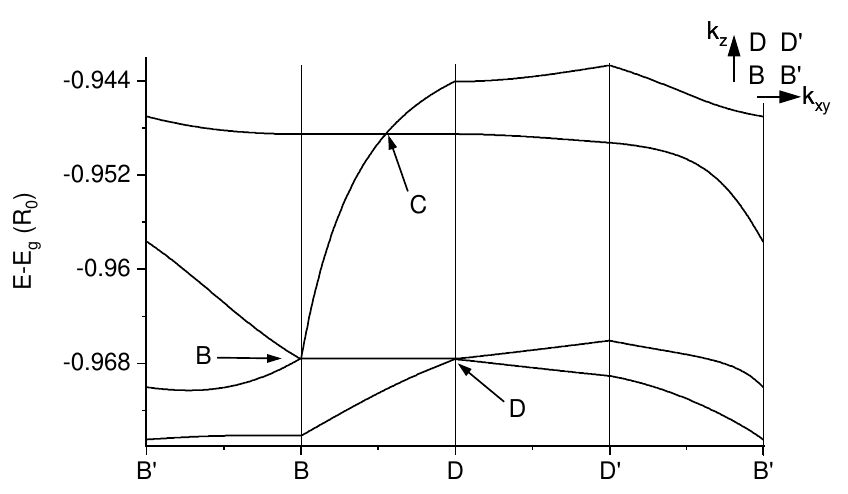}
\caption{Polariton dispersion relation for GaAs in a magnetic field, along a square path through $k$ space (horizontal axis). Two corners of the square are the Weyl points marked $B$ and $D$, here and on Fig.\ \ref{Fig:Weyl2}. The two remaining corners $B^\prime$ and $D^\prime$ are formed by displacing these points in a direction perpendicular to $k_z$, as indicated in the top right of the figure. There is a third Weyl point along one side of the square, marked $C$.\label{Fig:Weylbs}}
\end{figure}

Figure\ \ref{Fig:Weyl2} shows the dispersion relation near some of
these singularities. Since the dispersion has rotational symmetry around $z$, we plot it as a function of $k_z$ and the wavevector in the xy plane, $k_{xy}$. One of the degeneracies between the two circular
polarization modes seen in Fig.\ \ref{Fig:disps}(a) is point A in Fig.\ \ \ref{Fig:Weyl2}, lying between the orange and blue (second and third from the bottom) sheets. It can be seen to split linearly in $k_z$ but quadratically
in $k_{xy}$. The same structure appears for the other crossing of the
circular modes at $\theta=0$ (not shown). The Hamiltonian close to each of these degeneracies is therefore of the form 
\begin{align}
c_1(k_x^2 \sigma_x + k_y^2 \sigma_y) +c_2(k_z -k_c)\sigma_z
\end{align} for appropriate constants $c_1$ and $c_2$ and crossing 
point $k_c$. Following the procedure given in Ref. \cite{chang_chiral_2015}, we 
determine that such a dispersion implies these degeneracies have zero 
topological charge. However, Fig.\ \ref{Fig:Weyl2} also shows three 
crossings involving the z-polarized excitons, labeled B, C, and D. As can be seen in the figure, the dispersion near each of these points forms a pair of touching cones. The energy splitting is linear in all three components of the wave vector, measured from the degeneracy. This linear splitting is further shown in Fig.\ \ref{Fig:Weylbs}, and by the analysis below [Eq. (\ref{eq:weylpointlocalh})]. These points are, therefore, Weyl points. Their linear dispersion implies that they are topologically protected, carry a unit topological charge, and are monopoles of the Berry flux\ \cite{lu_weyl_2013,armitage_weyl_2018}. This is because the only form possible for a two-band Hamiltonian, with a degeneracy away from which the bands split linearly in all directions, is $\pm \bm{\sigma}.(\mathbf{k}-\mathbf{k}_c)$ (up to transformations of the coordinates). This implies topological protection, because any perturbations are additions of the Pauli matrices, which merely move the degeneracy. The unit topological charge can be confirmed using the approach in Ref.\ \cite{chang_chiral_2015} as above. The Weyl points B, C, and D correspond to three of the degeneracies between the transverse polaritons and the z-polarized excitons at $\theta=0$. The final fourth such degeneracy, lying in one of the longitudinal-transverse gaps\ \cite{YuCard}, is also a Weyl point, with a linear dispersion. However, it occurs at an imaginary $k$.

The dispersion perpendicular to the applied field ($\theta=\pi/2$) is
shown in Fig.~\ref{Fig:disps}(c). As indicated by the coloring, in this
case the transverse modes are purely linearly polarized, along the
polar vectors $\mathbf{e}_{\theta}=-\mathbf{z}$ and
$\mathbf{e}_{\phi}$. Considering this geometry, we identify two further
degeneracies in the polariton dispersion, where modes with these two
polarizations cross. These are both extended ring degeneracies, due to
the $\phi$ invariance of the system.

We see that there are, in total, eight distinct degeneracies of the
polariton dispersion relation in the region $0\leq \theta \leq
\pi/2$. Of these eight degeneracies, six are isolated degeneracies
occurring in the $\theta=0$ direction, and two are extended
degeneracies occurring in the $\theta=\pi/2$ plane. The six isolated
degeneracies divide into four Weyl points, one of which is at an
imaginary $k$, and two topologically trivial degeneracies with a mixed
quadratic-linear dispersion.

The Weyl points are degeneracies between the z-polarized excitons and
the xy-polarized polaritons at $\theta=0$. To see why this gives a
Weyl point, with a linear dispersion, we note that the coupling
between such modes --- and hence the splitting of the degeneracy --- is proportional to
$\sin(\theta)$ [see Eq. (\ref{eq:eightbyeight})]. This leads to a splitting of the degeneracy which is linear in the magnitude of the transverse wave vector, since $\sin(\theta)\approx\theta\propto |k_{xy}|$, where $k_{xy}=\pm\sqrt{k_x^2+k_y^2}$. Formally, the Hamiltonian for the two modes near the
degeneracy takes the form \begin{align}\begin{pmatrix} c^\prime
    (k_z-k_{c})+\omega_0 & \Omega \sin(\theta)/2 \\ \Omega
    \sin(\theta)/2 &
    \omega_x \end{pmatrix},\label{eq:weylpointlocalh}\end{align} which,
with $\sin(\theta)\approx k_x/k_c$ (for $k_y$=0), gives a linear
dispersion in $k_x$ and $k_z-k_c$ at the degeneracy
$\omega_0=\omega_x$. $\Omega$ is the strength of the coupling between
the z-polarized exciton and the polariton, involving the
amplitude, in the polariton, of the $\mathbf{e}_\theta$-polarized
photon. $k_c$ is the wave vector at the degeneracy,
$\omega_0=\omega_x=c^\prime k_c$ is the frequency, and $c^\prime$ is the
velocity. Notably, these Weyl points lie at the critical tilt between 
a type-I and a type-II point \cite{jalali-mola_electrodynamics_2019} and as such are the three-dimensional (Weyl) generalization of the 
recently achieved type-III Dirac point \cite{milicevic_type-iii_2019}. 
These classifications distinguish Weyl points based on their iso-frequency 
contours. In type-I cases the energy contour is a point, and in type-II cases 
the contours are surfaces, while type-III cases represent the transition 
between type-I and type-II cases where the contour is a line. 

\subsection{Exceptional points in the dispersion relation}

\begin{figure}
\includegraphics{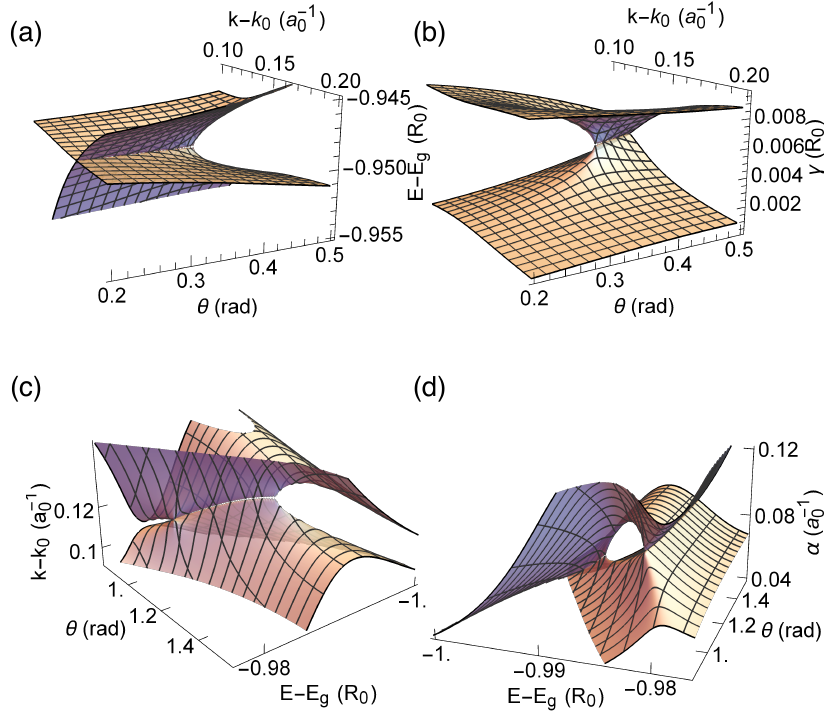}
\caption{Exceptional points in the polariton dispersion (top panels)
  and wave vector surface (bottom panels). (a) and (b) Real and imaginary
  parts of the polariton energies $E(k,\theta)$, for a real wavevector
  of magnitude $k$, at propagation direction $\theta$. Only one
  exciton, $X_3$, is damped, with rate $\gamma_3=0.01R_0/\hbar$. (c) and (d) Real and imaginary parts of the polariton wave vector
  $k(E,\theta)$, for a real energy $E$. All the excitons have an equal
  damping rate $\gamma=0.008R_0/\hbar$.\label{fig:epfigures}}
\end{figure}

In the presence of damping, the Hamiltonian, $H_8$, becomes
non-Hermitian, and the polariton dispersion, $\omega(k,\theta,\phi)$,
can contain rings of exceptional points arising from the Weyl points
described above. This can be seen by considering the local
Hamiltonian, Eq. (\ref{eq:weylpointlocalh}), for one of the Weyl
points. Damping will arise from the phonon or impurity scattering of
the polariton; exciton losses, through Auger recombination and
trapping in dark impurity states; and photon losses, from any
background absorption and the escape of light through the boundaries
of the sample. Such damping can be modeled by introducing imaginary
parts to the frequencies of the corresponding oscillators\
\cite{YuCard}, so that we replace
$\omega_{0,x}\rightarrow \omega_{0,x}-i\gamma_{0,x}$ in
Eq. (\ref{eq:weylpointlocalh}).

To understand the origins of the exceptional points, we consider the
eigenvalues of Eq. (\ref{eq:weylpointlocalh}) at the bare resonance,
which is achieved by tuning $k_z$ such that the real parts of the
diagonal elements are equal. At this point the splitting, in the
presence of damping, is
$\sqrt{\Omega^2\sin^2\theta-(\gamma_{0}-\gamma_{x})^2}$. Thus the imaginary
parts of the eigenvalues are split, and the real parts
degenerate, for angles less than
$\theta_c=\arcsin |\gamma_0-\gamma_x|/\Omega$, while the opposite is true
for angles greater than this value. This corresponds to the transition between the weak-coupling and strong-coupling regimes for the two modes\ \cite{kavokin_microcavities_2017}. In the weak-coupling regime the coupling strength $\Omega\sin\theta$ is smaller than the effective damping rate, and the normal modes are at the original degenerate frequencies, whereas in the strong-coupling regime it is larger, and the normal-mode frequencies (the real parts of the eigenvalues) are split. Because the
interaction between the modes depends on the angle we can use it to
access the transition between these two regimes. At the critical angle
$\theta_c$, marking the onset of strong-coupling between these modes, we see that
both the real and imaginary parts of the spectrum are degenerate, so
we have an exceptional point. Further details of the origins and
consequences of such exceptional points, which are generic features of
parameterized non-Hermitian eigenvalue problems, can be found in the
review by Heiss\ \cite{heiss_physics_2012}. In our case the degeneracy
occurs at all $\phi$, so that there is a ring of exceptional points in
the space of wavevector, where both the real and imaginary parts of the
polariton energies are degenerate and, as a corollary\
\cite{heiss_physics_2012}, the non-Hermitian Hamiltonian is defective.
  
In Fig.\ \ref{fig:epfigures} we show an exceptional point of this
type, in the spectrum of the full Hamiltonian, $H_8$. Figures\
\ref{fig:epfigures}(a) and \ref{fig:epfigures}(b) show the real and imaginary parts,
respectively, of the energies, and the expected local structure around
an exceptional point\ \cite{heiss_physics_2012} is clearly
visible. As can be seen from the analysis above, the angle of the
exceptional point depends on the difference in the damping constants; if this is too small, the Weyl exceptional ring will be indistinguishable from a Weyl point. For Fig.\ \ref{fig:epfigures} we have chosen to introduce damping $\gamma_3=0.01R_0/\hbar$ for the $X_3$ exciton only, so that the exceptional point structure is clear. This is a simple, nonrealistic choice, made to illustrate the exceptional points which occur more generally in $H_8$ when the modes have differing damping rates. Differences in the damping rates of the different exciton-polariton branches can be expected due to the differing dispersion relations and photon-exciton fractions, and from exciton spin relaxation\ \cite{maialle_exciton_1993, shelykh_spin_2005} among the Zeeman-split levels. 

\subsection{Exceptional points in the isofrequency surface}

\begin{figure}
\includegraphics{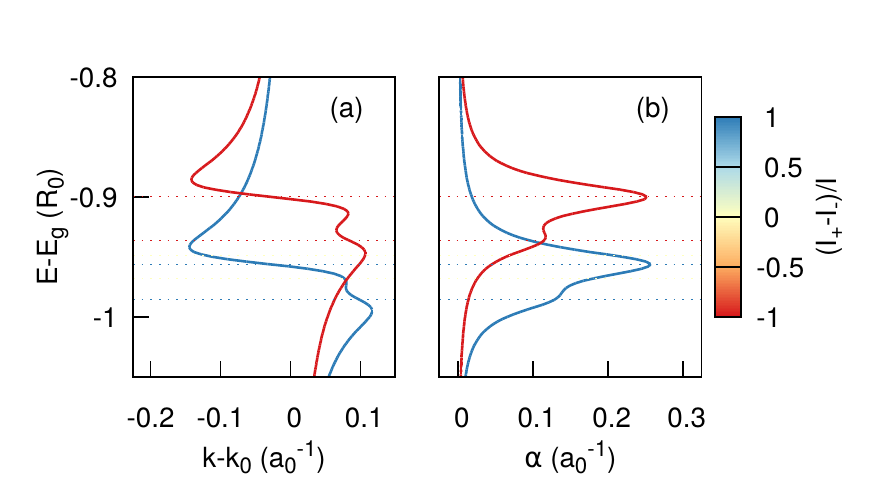}
\caption{(a) Real and (b) imaginary parts of the wave vector of the polariton dispersion relation at real energies, for propagation at $\theta=0$. Coloring shows the circular polarization of the modes. Dashed horizontal lines show the energies of the excitons colored by polarization. All excitons have an equal damping rate $\gamma=0.015R_0/\hbar$. }
\label{Fig:ZGpolaritons}
\end{figure}

We now consider the effect of damping in terms of the complex-valued
wave vector $k(E,\theta,\phi)=k(E,\theta)$ as a function of the
real-valued energy and propagation direction. This may be compared
with the treatment above, where we considered the complex-valued
energy as a function of the real-valued wave vector. At a particular
energy the two-sheeted function $k(E,\theta,\phi)$ is an isofrequency
surface of the dispersion relation, whose normals give the ray
directions\ \cite{BornWolf}.  More generally, real energies correspond to monochromatic
continuous-wave excitation, and the function $k(E,\theta)$ describes
the propagation of polaritons under such conditions. As we shall see,
the isofrequency surface can have rings of exceptional points (at
particular real energies), similar to those in the dispersion relation
(at particular real
wavevectors). 

Figure\ \ref{Fig:ZGpolaritons} shows the complex-valued wave vector
function obtained using the parameters of Fig.\ \ref{Fig:disps}(a) with
damping $\gamma=0.015 R_0/\hbar$ for the excitons. Here and throughout
this section, we consider an equal damping rate of all the excitons,
$\gamma_n=\gamma$. As can be seen, the damping blurs the distinction
between the lower, intermediate, and upper polariton branches, joining
them together for each polarization. In the imaginary parts of the
wave vectors, i.e., the absorption coefficients, we can see the
microstructure of the individual excitonic resonances and their
associated oscillator strengths. We see that there are energies where
the real parts of the wave vectors for the two polarizations are
degenerate, and there are, also, energies where the absorption
coefficients are degenerate.  These are not exceptional points,
however, as the degeneracies in the real and imaginary parts of $k$ occur
at different energies.

\begin{figure}
  \includegraphics{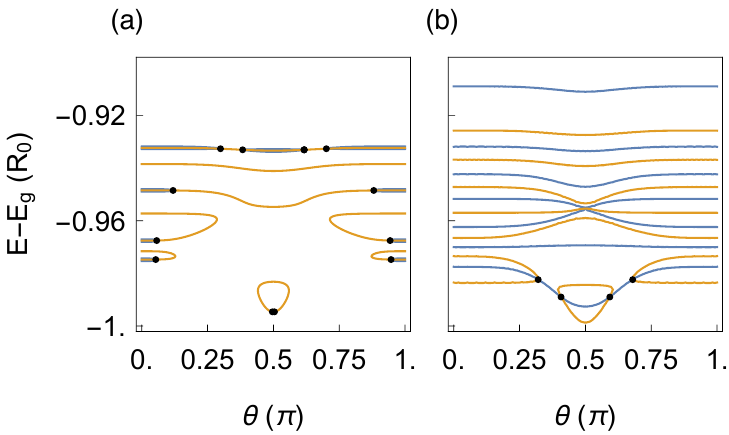}
  \caption{Exceptional points of the complex wavevector $k(E,\theta)$ for real energies, for two values of the exciton damping: $\gamma=0.0005R_0/\hbar$ (a), and $\gamma=0.008R_0/\hbar$ (b). The blue (orange) curves are the zero contours of the real (imaginary) part of the discriminant for the characteristic equation determining $k$. The points show the locations of the exceptional points.\label{Fig:DiscAnalysis}}
\end{figure}

The exceptional points of the wave vector function $k(E,\theta)$ are
values of $E$ and $\theta$ where, simultaneously, the real parts of
$k$ and the imaginary parts of $k$ are
degenerate. To identify these degeneracies, we consider the
characteristic equation for Eq.~(\ref{eq:twobytwomatrix}), which is a quadratic in $k$. The
exceptional points are the zeros of the discriminant of this
quadratic. They can be found by plotting the zero contours of its real
and imaginary parts and looking for their crossings. This is shown
for two different values of the damping rate in Fig.\
\ref{Fig:DiscAnalysis}.

Figure\ \ref{Fig:DiscAnalysis}(a) shows the situation for a small
damping rate, $\gamma=0.0005R_0/\hbar$. This corresponds to an exciton relaxation time $1/\gamma\approx 300$ ps, similar to that observed in quantum wells\ \cite{maialle_exciton_1993}. Figure\ \ref{Fig:DiscAnalysis}(b) shows the results for stronger damping,
$\gamma=0.008R_0/\hbar$, so that $1/\gamma\approx 20$ ps. In both cases we see that there are exceptional points,
which originate from the degeneracies of the polariton
dispersion in the absence of damping. As damping is introduced the
degeneracies of the polariton dispersion move in the $(\theta, E)$
plane. The richest structure in terms of degeneracies is at low
damping, as in Fig. \ref{Fig:DiscAnalysis}(a), where we see there are
six exceptional points in the region $0\leq \theta \leq \pi/2$.  Since
all of these points occur at non-zero $\theta$ they correspond to
rings of exceptional points in the isofrequency surface, at certain
energies, owing to the $\phi$ independence of the solutions. As the
damping is increased the exceptional points annihilate and their
number reduces, as can be seen in Fig.\ \ref{Fig:DiscAnalysis}(b), where
there are now only two exceptional points.

Figures~\ref{fig:epfigures}(c) and \ref{fig:epfigures}(d) show the real and imaginary parts
of the complex wave vector, as functions of the real energy $E$ and
angle $\theta$, in the region containing the two exceptional points of Fig.\ \ref{Fig:DiscAnalysis}(b). The two exceptional points are joined by a line degeneracy in the real parts of the wave vector, which is clearly visible in Fig.\ \ref{fig:epfigures}(c). The structure around each exceptional point may be compared with an exceptional
point in the complex energy [Figs.\ \ref{fig:epfigures}(a) and \ref{fig:epfigures}(b)]. We again have the expected general form, i.e., line
degeneracies in each of the real and imaginary parts, which meet at
the exceptional point. The overall structure may, also, be compared with that described by Berry and Dennis \cite{BerryBr1} for frequency-independent absorbing dielectrics, for which the complex function $k(\theta,\phi)$ contains degeneracies in particular wavevector directions. These exceptional points define the singular axes of the crystal. They are points in the space of wave vector direction, but occur at all frequencies. The degeneracies of the complex-valued wave vector described here are, instead, extended in the space of wave vector direction (forming rings), but occur only at specific frequencies. 

\section{Conclusion}

The strong coupling of light to excitons in a magnetic field gives
rise to topologically nontrivial dispersion relations $\omega(k)$,
and wave vector surfaces $k(E)$, for polaritons in bulk zinc-blende
semiconductors. The complex degeneracy structure of the dispersion
provides a route to realizing topological effects for polaritons in
three dimensions, going beyond previous work in two-dimensional \cite{klembt_exciton-polariton_2018,ge_floquet_2018,PolZG,PolZ,li_lieb_2018,kartashov_two-dimensional_2019,YiKar,TopPol}, and also one-dimensional \cite{solnyshkov_kibble-zurek_2016,downing_topological_2019},
systems such as microcavities. In the absence of dissipation the polariton
dispersion contains Weyl points, for propagation along the field, and
ring degeneracies, for propagation perpendicular to it. In the
presence of dissipation the Weyl points become rings of exceptional
points, which generalize the corresponding Dirac exceptional rings of
two-dimensional dissipative systems \cite{EPRing}. A realization of
Weyl exceptional rings in cold atomic gases has recently been
proposed \cite{xu_weyl_2017}; the present work shows that a different
realization in semiconductors may be possible.

Topological bands, Weyl points, and surface states (Fermi arcs) have
recently been revealed in transmission experiments
\cite{ExpObsWeyl,HalfPol,yang_ideal_2018,chen_photonic_2016,WeylObs2}
on photonic bandstructures \cite{lu_weyl_2013}. The topological
dispersion relations described here, and their consequences, would
also give signatures in transmission. Polariton spectra are, however, typically determined using angle-dependent reflectivity measurements. Such techniques are well established~\cite{sell_polariton_1973}, and have been used to determine exciton-polariton spectra in bulk semiconductors, with resolution sufficient to resolve
Zeeman-split
levels~\cite{willmann_magneto-reflectance_1974,
  nam_free-exciton_1976,
  nam_free-exciton_1976-1,chen_exciton-polaritons_1987, dingle_magneto-optical_1973}. The exciton-polariton dispersion gives rise to non-standard lineshapes in reflectivity, so the spectra need to be interpreted by comparison to coupled oscillator models\ \cite{sell_polariton_1973}. Another approach would be to pump incoherently above the polariton branches, and study the spectra in photoluminescence. This 
creates the further possibility of exploring the impact of gain on the
topological bands and surface modes, and creating a polaritonic
topological laser based on surface states. Perhaps the most promising
way, however, in which the polariton system goes beyond existing
photonic topological materials is through the presence of large
nonlinearities, giving it potential for realizing topological
strong-interaction effects using light.

An important question is the extent to which the effects described
here can be observed, given realistic values of the dissipation. We
see from Figs.~\ref{Fig:1sexcitons} and~\ref{Fig:Weyl2} that the
scale of the multiple polariton branches is set by the Zeeman
splittings in the magnetoexciton spectrum. Note that the light-matter
coupling strength, $\hbar\Omega_0$, is much larger, and therefore not
a limiting factor. To observe the multiple polariton bands, along with
the Weyl points, will require the exciton damping to be smaller than
at least some, or ideally all, of the splittings. The achievability of such a regime is demonstrated by
many reflectivity and photoluminescence experiments on zinc-blende
semiconductors, in which polaritons formed from Zeeman-split excitons
are resolved\ \cite{willmann_magneto-reflectance_1974,
  nam_free-exciton_1976,
  nam_free-exciton_1976-1,chen_exciton-polaritons_1987,dingle_magneto-optical_1973}. In those works some, but not all, of the Zeeman lines are resolved at the relatively low field $B_{\mathrm{max}}$ considered here. However, all of the lines are resolved at higher fields, where the splittings become larger. Although such fields invalidate the perturbative calculation of the exciton energies we took as input to our model, the exciton levels and their polarizations remain. Thus the qualitative features of the polariton spectra predicted by our model will survive, with only quantitative changes.

\section*{Acknowledgments}

We acknowledge support from the Irish Research Council through Award No. GOIPG/2015/3570 and the Science Foundation Ireland through Award No. 15/IACA/3402.

\appendix*
\section{Polariton Hamiltonian matrix}

As discussed in the main text, we obtain the polariton spectrum from the Heisenberg equations of motion for the exciton and photon annihilation operators. In terms of the vector of annihilation operators \begin{displaymath}\psi=(\hat{a}_+,\hat{a}_-,\hat{x}_2,\hat{x}_3,\ldots,\hat{x}_7)^{T}\end{displaymath} this reads \begin{equation}i\hbar\frac{d\psi}{dt}=H_8\psi=\hbar\omega\psi,\label{eq:heisnberg} \end{equation} where the second equality holds in an eigenstate, in which all the annihilation operators have time dependence $e^{-i\omega t}$. $\hat{a}_{+,-}$ correspond to the two circularly polarized photon modes, and $\hat{x}_{2\ldots7}$ correspond to the six optically active exciton modes. The wave vector labels are suppressed for brevity. The explicit form of the matrix $H_8$ is
\begin{widetext}
\begin{align}
&H_8/\hbar=\nonumber\\
&\begin{pmatrix}
\frac{ c k}{\sqrt{\epsilon}} & 0 & \hdotsfor{6} \\ 
0 &  \frac{  c k}{\sqrt{\epsilon}} & \hdotsfor{6} \\ 
 i\frac{  \Omega_0}{4\sqrt{3}}c_+(\theta)e^{-i\phi} &-i\frac{  \Omega_0}{4\sqrt{3}}c_-(\theta)e^{-i\phi}  & \omega_2 - i \gamma_2 & 0 & 0 & 0 & 0 & 0\\
  -i\frac{  \Omega_0}{2\sqrt{3}} \sin \theta &-i\frac{  \Omega_0}{2\sqrt{3}} \sin \theta  & 0 & \omega_3 - i \gamma_3 & 0 & 0 & 0 & 0\\
 i\frac{  \Omega_0}{4} c_-(\theta)e^{i\phi} &-i\frac{  \Omega_0}{4}c_+(\theta)e^{i\phi}  & 0 & 0 & \omega_4 - i \gamma_4  & 0 & 0 & 0 \\
 i\frac{  \Omega_0}{4}c_+(\theta)e^{-i\phi} &-i\frac{  \Omega_0}{4}c_-(\theta)e^{-i\phi}  & 0 & 0 & 0 & \omega_5 - i \gamma_5  & 0 & 0  \\
-i\frac{  \Omega_0}{2\sqrt{3}} \sin \theta &-i\frac{  \Omega_0}{2\sqrt{3}} \sin \theta & 0 & 0 & 0 &0 & \omega_6 - i \gamma_6  &  0 \\
i\frac{  \Omega_0}{4\sqrt{3}}c_-(\theta)e^{i\phi} & -i\frac{  \Omega_0}{4\sqrt{3}}c_+(\theta)e^{i\phi}    & 0 & 0 & 0 &0 &  0 & \omega_7 - i \gamma_7 
\end{pmatrix}\label{eq:eightbyeight}
\end{align}
\end{widetext}
where the omitted parts of the first two rows are Hermitian conjugates of
the corresponding parts of the first two columns,
$c_{\pm}(\theta) = (1\pm \cos\theta)$, and
$\omega_{2\ldots 7}=E_{2\ldots7}/\hbar$ are the exciton
frequencies. We treat damping by taking the expectation value of
Eq. (\ref{eq:heisnberg}), so that it becomes a set of coupled
equations for the complex amplitudes, in which we can introduce
imaginary parts $\gamma_{2\ldots 7}$ in the exciton frequencies as
indicated. The now complex polariton frequencies $\omega$ are then
computed as the eigenvalues of $H_8/\hbar$.


%

\end{document}